\documentclass[a4paper]{article}
\pdfoutput=1
\usepackage{amsmath,amssymb}
\usepackage{algorithm}
\usepackage{algpseudocode}
\usepackage{pgfplots}
\usepackage{enumerate}
\usepackage{url}
\usepackage{authblk}
\usepackage[numbers]{natbib}

\newcommand{\dd}{\mathrm{d}}
\newcommand{\derfrac}[2]{\frac{\dd #1}{\dd #2}}
\newcommand{\pderfrac}[2]{\frac{\partial #1}{\partial #2}}
\DeclareMathOperator{\cd}{|}
\DeclareMathOperator*{\argmax}{argmax}

\begin{document}

\title{A Fast Algorithm for Sampling from the Posterior of a von Mises distribution\footnote{This is an author-created, un-copyedited version of an article accepted for publication in the \emph{Journal of Statistical Computation and Simulation}. The Version of Record is available online at http://dx.doi.org/doi:10.1080/00949655.2014.928711.}}

\author{Peter G.M. Forbes$^{a}$\thanks{Corresponding author. Email: forbes@stats.ox.ac.uk\vspace{6pt}}  
\, and Kanti V. Mardia$^{b,a}$\\
\vspace{6pt}
$^{a}${Department of Statistics, University of Oxford,\\ 1 South Parks Road, Oxford OX1 3TG, UK}\\
$^{b}${Department of Statistics, School of Mathematics,\\ University of Leeds, Leeds LS2 9JT, UK}}

\maketitle

\begin{abstract}
Motivated by molecular biology, there has been an upsurge of research activities in directional statistics in general and its Bayesian aspect in particular. The central distribution for the circular case is von Mises distribution which has two parameters (mean and concentration) akin to the univariate normal distribution. However, there has been a challenge to  sample efficiently from the posterior distribution of the concentration parameter. We describe a novel, highly efficient algorithm to sample from the posterior distribution and fill this long-standing gap.
\end{abstract}

\section{Introduction}
\label{sect:intro}
There has been renewed interest in directional Bayesian analysis in view of its fundamental applications to molecular biology \cite{boomsma,frellsen,mardia2013}. Due to chemical constraints on the bonds of biomolecules, the geometry of these molecules can be described by a set of angles. Other applications include locating and tracking an electric signal \cite{guttorp} and the analysis of forensic fingerprint evidence \cite{forbeslasr}.  All these applications involve circular data which is naturally modelled by the von Mises distribution. 

The probability density function of the von Mises distribution with mean $\mu\in S^1$ on the unit circle and concentration parameter $\kappa\geq0$ is given by \cite{mardiabook}
\[
p(\theta)=\frac{1}{2\pi I_0(\kappa)}\exp\left\{\kappa\cos(\theta-\mu)\right\},
\]
where $I_m(\cdot)$ is the modified Bessel function of the first kind and order $m$.  The circular variance can be described by $1-r(\kappa)$ where $r(\kappa)=I_1(\kappa)/I_0(\kappa)$. Let $\boldsymbol{\theta}=(\theta_1,\ldots,\theta_n)$ be a vector of observations from a von Mises distribution. When a conjugate prior is used, the posterior distribution of the mean $\mu\cd \boldsymbol{\theta},\kappa$ is itself von Mises, and can be easily sampled via \cite{bestfisher}.

Let $\pi(\kappa)\propto I_0(\kappa)^{-a}\exp(-b\kappa)$ be the conjugate prior for the concentration.  The posterior is
\begin{equation}
p(\kappa)=\frac{A}{I_0(\kappa)^\eta} \exp(-\eta\beta_0\kappa)
\label{eq:basicdensity}
\end{equation}
where $\eta>0$ and $\beta_0\in (-1,\infty)$ are observed constants;  in this case $\eta=a+n$ and $\beta_0=b/(a+n)-n^{-1}\sum_{i=1}^n \cos(\theta_i-\mu)$.   For the case $\eta=1$ and $\beta_0>1$ the normalization constant is $A=\sqrt{\beta_0^2-1}$, however in general the normalization constant is intractable \cite{mardialasr}.  In this paper we  shall call \eqref{eq:basicdensity} the \emph{Bessel exponential distribution}.

Existing algorithms to sample \eqref{eq:basicdensity} tend to generate from approximate distributions \cite{guttorp} or have a large overhead of sampled auxiliary variables \cite{damien}.  We present a new, extremely fast algorithm to sample from the Bessel exponential distribution.
 
For large $\kappa$, $I_0(\kappa)$ is approximately $\exp(\kappa)/\sqrt{2\pi\kappa}$ \cite[eq. 9.7.1]{specialfcnshandbook}.  Plugging this approximation into \eqref{eq:basicdensity} yields a gamma density with shape $\eta/2+1$ and rate $\eta(\beta_0-1)$.

This insight  motivates us to use a gamma-based acceptance-rejection sampler for $\kappa$.  However, the above approximation for $I_0(\kappa)$ breaks down for small $\kappa$, and thus great care is needed to ensure our rejection sampler is efficient for all values of $\kappa$.  We derive the optimal gamma-based proposal distribution and show that the resulting sampler has an acceptance probability of at least 0.7 for all $\eta$ and $\beta_0$.  The minimum acceptance probability of $\approx 0.7$ occurs when the distribution is concentrated around $\kappa=0$. 

The algorithm is described in Section~\ref{sect:alg} and derived in Section~\ref{sect:derivation}.  Enhancements are considered in Section~\ref{sect:speedup} and the algorithm's efficiency is explored in Section~\ref{sect:efficiency}.

\section{The algorithm}
\label{sect:alg}

As discussed above, we can approximate the Bessel exponential distribution with a gamma distribution.  However, because the ratio of these densities diverges as $\kappa\rightarrow 0$, we cannot directly use a gamma proposal for our rejection sampler. Instead we propose values $\kappa=x-\varepsilon$ where $\varepsilon>0$ and $x$ has a gamma distribution.  This is an application of Marsaglia's exact approximation procedure, see \cite{exactApprox} for more details.

Using a shifted gamma proposal with shape $\eta\alpha+1\geq 1$ and scale $\eta\beta>0$ leads to the envelope function
\[
q(\kappa;\alpha,\beta,\varepsilon)=M(\alpha,\beta,\varepsilon)(\kappa+\varepsilon)^{\eta\alpha}\exp(-\eta\beta\kappa);\quad\kappa\geq 0
\]
for $p(\kappa)$, where the amplitude $M(\alpha,\beta,\varepsilon)$ is chosen to ensure the ratio $p/q$ is bounded below one. We can generate a sample from the Bessel exponential distribution by generating a sample $\kappa$ from $q$ and accepting it with probability
\[
\frac{p(\kappa)}{q(\kappa;\alpha,\beta,\varepsilon)}=\frac{A}{M(\alpha,\beta,\varepsilon)}\exp\{\eta g(\kappa;\alpha,\beta,\varepsilon)\}
\]
where
\begin{align*}
g(\kappa;\alpha,\beta,\varepsilon)&=(\beta-\beta_0) \kappa-\alpha\log (\kappa+\varepsilon)-\log I_0(\kappa),\\
M(\alpha,\beta,\varepsilon)&=A\exp\{\eta g(\kappa_0;\alpha,\beta,\varepsilon)\}
\end{align*}
and $\kappa_0=\argmax_{\kappa\geq 0}g(\kappa;\alpha,\beta,\varepsilon)$.  The samples generated via this procedure have the correct Bessel exponential distribution for any choice of the proposal parameters $\alpha,\beta,\varepsilon$, though the values of these parameters will affect the algorithm's efficiency.

In Section~\ref{sect:derivation} we show that the approximate optimal choices for the proposal parameters are
\begin{gather*}
\beta=\begin{cases}
\beta_0+1 \quad&\mbox{if }\beta_0\leq 1/(4\eta)-2/\left(3\sqrt{\eta}\right)\\
\beta_0+r(\kappa_0)+\dfrac{1-r(\kappa_0)}{1+40\eta\{\beta_0-1/(4\eta)+2/\left(3\sqrt{\eta}\right)\}^2}&\mbox{otherwise,}
\end{cases}\\
\varepsilon=\frac{\kappa_0\mathcal W_0\{c_3\exp(c_3)\}}{c_3-\mathcal W_0\{c_3\exp(c_3)\}},\quad\alpha=\{\beta-\beta_0-r(\kappa_0)\}(\kappa_0+\varepsilon)
\end{gather*}
where the terms $\kappa_0$ and $c_3$ are
\begin{gather*}
\kappa_0=\dfrac{1-1/{\eta}+1/(2\eta^2)}{\eta\beta_0+\sqrt{2\eta+\eta^2\beta_0^2}}+
\dfrac{1+3/(2\eta)-1/(4\eta^3)}{(\eta+1)\beta_0+\sqrt{2\eta+1+\eta^2\beta_0^2}},\\
\quad c_3=-\frac{\beta-\beta_0-\log\{I_0(\kappa_0)\}/\kappa_0}{\beta-\beta_0-r(\kappa_0)},
\end{gather*}
and $\mathcal W_0(\cdot)$ is the principal branch of the Lambert W function defined as $t=\mathcal W_0(t)\exp\{\mathcal W_0(t)\}$ for $\mathcal W_0(t)>-1$.

The acceptance-rejection algorithm to generate a sample from the Bessel exponential distribution proceeds as follows:
\begin{enumerate}[1.]
  \item Find efficient proposal parameters $\alpha,\beta,\varepsilon$, which will depend on $\eta,\beta_0$.
  \item Draw $x$ from a gamma distribution with shape $\eta\alpha+1$ and rate $\eta\beta$.
  \item Draw $u$ from a Uniform distribution on $[0,1]$.
  \item Accept $\kappa=x-\varepsilon$ if $\log u <\eta g(\kappa;\alpha,\beta,\varepsilon)-\eta g(\kappa_0;\alpha,\beta,\varepsilon)$, else go to 2.
\end{enumerate}
The detailed procedure is described in Algorithm~\ref{alg:kappa}.  When implementing the algorithm, both the Bessel functions and the Lambert W function can be computed using software such as the General Scientific Library \cite{gsl} or its R wrapper, the CRAN package \textbf{gsl}.  In practice it is often possible to avoid computing these functions, as we show in Section~\ref{sect:speedup}.

\begin{algorithm}[htb]
\begin{algorithmic}[1]
\State\Comment{Initialization: find parameters for proposal distribution}
\State $\kappa_L\gets 2/\left(\eta\beta_0+\sqrt{2\eta+\eta^2\beta_0^2}\right)$ \label{algline:startsetup}
\State $\kappa_U\gets(2+1/\eta)/\left\{(\eta+1)\beta_0+\sqrt{2\eta+1+\eta^2\beta_0^2}\right\}$
\State $c_1=1/2+\{1-1/(2\eta)\}/2\eta$
\State $\kappa_0\gets (1-c_1)\kappa_L+c_1\kappa_U$\label{algline:kappastar}
\State $i_0\gets I_0(\kappa_0)$
\State $r\gets I_1(\kappa_0)/i_0$
\State $c_2\gets 1/(4\eta)-2/\left(3\sqrt{\eta}\right)$
\If{$\beta_0\leq c_2$}
\State $\beta\gets \beta_0+1$
\Else
\State $\beta\gets\beta_0+r+(1-r)/\{1+40\eta(\beta_0-c_2)^2\}$ \label{algline:betadefn}
\EndIf
\State $c_3\gets \{\log(i_0)/\kappa_0-\beta+\beta_0\}/(\beta-\beta_0-r)$
\State $c_4\gets\mathcal W_0\{c_3\exp(c_3)\}$\label{algline:lambert}\Comment{Lambert's W function, see Section~\ref{sect:speedup}}
\State $\varepsilon\gets c_4\kappa_0/(c_3-c_4)$
\State $\alpha\gets (\beta-\beta_0-r)(\kappa_0+\varepsilon)$ \label{algline:endsetup}
\State\Comment{Perform rejection sampling}
\Repeat\label{algline:startloop}
\State $x\gets$ sample from a $\mathrm{Gamma}(\eta\alpha+1,\eta\beta)$ left-truncated at $\varepsilon$ \label{algline:xsampling}
\State $\kappa\gets x-\varepsilon$
\State $u\gets$ sample from a $\mathrm{Uniform}(0,1)$
\Until{$\log(u)/\eta<(\beta-\beta_0)(\kappa-\kappa_0)-\alpha\log
\left\{(\kappa+\varepsilon)/(\kappa_0+\varepsilon)\right\}
-\log\{I_0(\kappa)/i_0\}$}\label{algline:acceptreject}
\\\Return{$\kappa$}\label{algline:returnline}
\end{algorithmic}
\caption{Rejection sampler for the Bessel exponential distribution}
\label{alg:kappa}
\end{algorithm}

\section{Derivation of the algorithm}
\label{sect:derivation}

We will now derive the optimal parameters $\alpha,\beta,\varepsilon$ for the proposal distribution of $x=\kappa-\varepsilon$ which follows a gamma distribution with shape $\eta\alpha+1$ and rate $\eta\beta$.  We do so by maximizing the expected probability of acceptance
\begin{align}
E_{x\cd \alpha,\beta}&[I(x\geq\varepsilon)\exp\{\eta g(x-\varepsilon;\alpha,\beta,\varepsilon)-\eta g(\kappa_0;\alpha,\beta,\varepsilon)\}]\nonumber
\\&=\frac{(\eta \beta)^{\eta \alpha+1}}{\Gamma(\eta \alpha+1)}\exp\{-\eta \beta\varepsilon-\eta g(\kappa_0;\alpha,\beta,\varepsilon)\}\int_{0}^\infty \frac{\exp(-\eta \beta_0\kappa)}{I_0(\kappa)^\eta }\dd\kappa
\label{eq:acceptprob}
\end{align}
over $\alpha,\beta,\varepsilon,\kappa_0$ subject to the constraint $\kappa_0=\argmax_{\kappa\geq 0}g(\kappa;\alpha,\beta,\varepsilon)$. In order for the maximum to be finite as $\kappa\rightarrow\infty$ we require $\beta\leq \beta_0+1$.

By taking logs we see that maximizing \eqref{eq:acceptprob} with respect to $\kappa$ is equivalent to maximizing
\begin{equation}
h(\kappa_0;\alpha,\beta,\varepsilon)=(\alpha+\eta ^{-1})\log(\eta \beta)-\eta ^{-1}\log\Gamma(\eta \alpha+1)-\beta\varepsilon-g(\kappa_0;\alpha,\beta,\varepsilon).
\label{eq:hfun}
\end{equation}

The constraint $\kappa_0=\argmax_{\kappa\geq 0}g(\kappa;\alpha,\beta,\varepsilon)$ implies either $\kappa_0=0$ or $\frac{\dd}{\dd\kappa}g(\kappa=\kappa_0;\alpha,\beta,\varepsilon)=0$.  The Lagrangians for constrained optimization corresponding to these conditions are $h+\lambda\kappa_0$ and $h+\lambda\frac{\dd}{\dd\kappa}g$, neither of which have interior critical points over $(\alpha,\beta,\varepsilon,\lambda)$ because the $\alpha$ and $\varepsilon$ derivatives have no common root.  Thus the optimal parameters must lie on the boundary of the parameter space.  An examination of the boundaries show that the maximum satisfies $\frac{\dd}{\dd\kappa}g(\kappa=\kappa_0;\alpha,\beta,\varepsilon)=0$ and $g(\kappa_0;\alpha,\beta,\varepsilon)=g(0;\alpha,\beta,\varepsilon)$. Intuitively, this says that for any $\kappa_0$ we should pick $\varepsilon$ as small as possible while still having $\kappa_0$ be the maximizer of $g$. Thus our Lagrangian is
\begin{multline*}
L(\kappa_0;\alpha,\beta,\varepsilon,\lambda_1,\lambda_2)=h(\kappa_0;\alpha,\beta,\varepsilon)+\lambda_1\derfrac{g(\kappa=\kappa_0;\alpha,\beta,\varepsilon)}{\kappa}
\\+\lambda_2\{g(\kappa_0;\alpha,\beta,\varepsilon)-g(0;\alpha,\beta,\varepsilon)\}.
\label{eq:Lagrangian}
\end{multline*}

Our optimal parameters are either a critical point of $L$ or lie on one or more of the boundaries $\alpha=0,\varepsilon=0$ or $\beta=\beta_0+1$.  If either $\alpha=0$ or $\varepsilon=0$ then direct differentiation shows the maximum occurs when $\alpha=0,\varepsilon=0,\beta=\beta_0+r(\kappa)$ and $\kappa_0$ is the unique positive root of $\beta-1/(\eta\kappa)$.  We shall see this is a limiting case of the critical point solution.  The only other boundary is $\beta=\beta_0+1$; we shall see this is the solution when $\beta_0$ is close to $-1$.

To find the critical points of $L$, we start by setting the derivatives with respect to $\alpha,\varepsilon,\lambda_1$ and $\lambda_2$ to zero and rearranging yields the optimal parameters as functions of $\kappa_0$ and $\beta$. This yields
\begin{gather}
\alpha=\{\beta-\beta_0-r(\kappa_0)\}(\kappa_0+\varepsilon)\nonumber,\\
\varepsilon=\frac{\kappa_0\mathcal W_0\{c_3\exp(c_3)\}}{c_3-\mathcal W_0\{c_3\exp(c_3)\}},\quad c_3=-\frac{\beta-\beta_0-\log\{I_0(\kappa_0)\}/\kappa_0}{\beta-\beta_0-r(\kappa_0)}
\label{eq:params_beta_k},\\
\lambda_2=\frac{\Psi(\eta \alpha+1)-\log\{\eta \beta(\kappa_0+\varepsilon)\}-1+\beta(\kappa_0+\varepsilon)/\alpha}{\log(1+\kappa_0/\varepsilon)-\kappa_0/\varepsilon},\nonumber\\
\lambda_1=\frac{\kappa_0+\varepsilon}{\alpha\varepsilon}\{\beta\varepsilon^2+(\kappa_0\beta-\alpha)\varepsilon+\alpha\lambda_2\kappa_0\},\nonumber
\end{gather}
where $\Psi(x)=\derfrac{}{x}\log\Gamma(x)$ is the digamma function.  We must have $\beta>\beta_0+r(\kappa_0)$ so that $\alpha$ and $\varepsilon$ are positive. Notice that the limit $\beta\rightarrow \beta_0+r(\kappa_0)$ corresponds to the boundary case $\alpha=\varepsilon=0$ discussed above.

The above equations give all optimal parameters in terms of $\beta$ and $\kappa_0$.  Note that since constraint $\frac{\dd}{\dd\kappa} g(\kappa;\alpha,\beta,\varepsilon)=0$ at $\kappa=\kappa_0$ is satisfied whenever $\alpha=\{\beta-\beta_0-r(\kappa_0)\}(\kappa_0+\varepsilon)$, we are free to choose alternative, sub-optimal values for the other parameters if the true optimal values are too difficult to compute.  We shall explore this in Section~\ref{sect:speedup} when we use an approximation for the Lambert W function.

Finally, we find the optimal $\beta$ as follows.  This value must either lie on the boundary $\beta=\beta_0+1$ or else satisfy
\begin{equation}
\pderfrac{L(\kappa_0;\alpha,\beta,\varepsilon,\lambda_1,\lambda_2)}{\beta}=(\alpha+1/\eta )/\beta-(\kappa_0+\varepsilon)+\lambda_1-\lambda_2\kappa_0=0.
\label{eq:betaeq}
\end{equation}
Unfortunately plugging \eqref{eq:params_beta_k} into \eqref{eq:betaeq} and solving for $\beta$ as a function of $\kappa_0$ alone is analytically intractable.  However, one can check that $\partial L/\partial\beta$ decreases from positive infinity at $\beta=\beta_0+r(\kappa_0)$ to negative infinity as $\beta\rightarrow\infty$.  Since all admissible $\beta$ lie in the finite interval
\[\max\{0,\beta_0+r(\kappa_0)\}<\beta<\beta_0+1,\]
we can easily find the optimal $\beta$ through any standard one-dimensional root-finding algorithm.  If the root lies to the right of $\beta_0+1$, the optimal value is $\beta=\beta_0+1$.  

We plug the optimal $\beta$ into \eqref{eq:params_beta_k} to find all of our optimal parameters in terms of $\kappa_0$.  Doing this for each $\kappa_0$ and plugging the resulting parameters $(\alpha,\beta,\varepsilon)$ into $h(\kappa_0;\alpha,\beta,\varepsilon)$ yields a function which we numerically maximize over $\kappa_0$. Let $\kappa^*$ be the optimal value of $\kappa_0$ and let $(\alpha^*,\beta^*,\varepsilon^*)$ be the optimal parameters corresponding to $\kappa^*$. These are the desired parameters that maximize the expected acceptance probability \eqref{eq:acceptprob}.

The above numeric maximizations for $\beta$ and $\kappa_0$ may be acceptable when $\eta$ and $\beta_0$ are known a priori.  However, they are computationally prohibitive in the standard Monte Carlo case where we wish to generate many samples from the Bessel exponential distribution with different values of $\eta$ and $\beta_0$ for each sample. Thus our next task is to approximate the optimal parameters with easily computable functions of $\eta$ and $\beta_0$.

For all $\eta$ and $\beta_0$, $\kappa^*$ is well approximated by $\kappa_a$, the positive root of $\beta_0+r(\kappa)-1/(\eta\kappa)$; indeed $\kappa_a$ is the exact optimum in the boundary case $\alpha=\varepsilon=0$.  To approximate $\kappa_a$ we use the bounds \cite[eq. 11]{amos},
\[ \frac{\kappa}{1+\sqrt{1+\kappa^2}}\leq r(\kappa) \leq \frac{\kappa}{\sqrt{4+\kappa^2}}.\]
Rearranging these bounds shows that $\kappa_L\leq\kappa_a\leq \kappa_U$ where 
\[\kappa_L=2\left(\eta \beta_0+\sqrt{2\eta +\eta ^2\beta_0^2}\right)^{-1},\quad \kappa_U=(2+1/\eta )\left\{(\eta +1)\beta_0+\sqrt{2\eta +1+\eta ^2\beta_0^2}\right\}^{-1}.\]
These bounds are relatively tight, we found that the convex combination $(1-c_1)\kappa_L+c_1\kappa_U$ with $c_1=1/2+\{1-1/(2\eta )\}/2\eta $ provides a good approximation to $\kappa_a$ and hence to $\kappa^*$.

The parameter $\beta^*$ is exactly equal to $\beta_0+1$ when $\beta_0$ is close to its lower bound of negative one.  For $\beta_0$ sufficiently large, $\beta^*$ drops from its upper limit $\beta_0+1$ towards its lower limit $\beta_0+r(\kappa^*)$.  The transition between the two limits is very rapid for $\eta>10$.  We achieve good accuracy with the approximation
\begin{equation*}
\beta^*\approx
\begin{cases}
\beta_0+1 \quad & \mbox{if } \beta_0<c_2 \\
\beta_0+r(\kappa^*)+\dfrac{1-r(\kappa^*)}{1+40\eta (\beta_0-c_2)^2} & \mbox{otherwise},
\end{cases}
\end{equation*}
where $c_2=1/(4\eta )-2/\left(3\sqrt{\eta }\right)$.  This approximation is very good when $\eta $ is large or when $|\beta_0|>0.1$.  A slower, more precise approximation for $\beta^*$ may lead to parameters which provide better efficiency for small $\eta$ and $\beta_0\approx 0$; we address this in Section~\ref{sect:efficiency}. 

Given these approximations of $\kappa^*$ and $\beta^*$, the parameters $\alpha^*$ and $\varepsilon^*$ are given by \eqref{eq:params_beta_k}.

\section{Further speed enhancements}
\label{sect:speedup}

The truncated gamma on line~\ref{algline:xsampling} can be sampled using Dagpunar's algorithm \cite{dagpunar}.  Alternatively, one can use a standard gamma sampling algorithm such as Marsaglia--Tsang \cite{gammasampler} and reject when $x<\varepsilon$. Indeed, the Marsaglia--Tsang algorithm is itself a rejection sampler with a Gaussian proposal, and its rejection step can be combined with the rejection step on line~\ref{algline:acceptreject} for an additional speed-up.

The function $\mathcal W_0$ on line~\ref{algline:lambert} can be approximated by (Winitzki, \cite{winitzki})
\begin{equation}
\mathcal W_0(t)=\frac{et}{1+\{(2et+2)^{-1/2}+(e-1)^{-1}-2^{-1/2}\}^{-1}}
\label{eq:winitzki}
\end{equation}
with no noticeable drop in the expected probability of acceptance.

Finally, we can implement the simple squeezes
\begin{align*}
I_0(\kappa) &< \{1+1/(2\kappa)\}\exp(\kappa)/\sqrt{2\pi\kappa}:\quad&\kappa>0 \\
I_0(\kappa) &> \exp(\kappa)/\sqrt{2\pi\kappa}:\quad&\kappa>0.259,
\end{align*}
to avoid computing the Bessel function within the rejection loop on line~\ref{algline:acceptreject}. Specifically, the loop on lines \ref{algline:startloop} to \ref{algline:returnline} can be replaced with Algorithm~\ref{alg:squeezed}.
\begin{algorithm}[htb]
\begin{algorithmic}
\State $c_5 = \log(i_0)$
\Loop
\State $x\gets$ sample from a $\mathrm{Gamma}(\eta \alpha+1,\eta \beta)$ left-truncated at $\varepsilon$
\State $\kappa\gets x-\varepsilon$
\State $c_6\gets\frac{1}{2}\log(2\pi\kappa)-\kappa$
\State $u\gets$ sample from a $\mathrm{Uniform}(0,1)$
\State $v\gets \log(u)/\eta -(\beta-\beta_0)(\kappa-\kappa_0)+\alpha\log\{(\kappa+\varepsilon)/(\kappa_0+\varepsilon)\}-c_5$
\State\textbf{if } $\kappa< 0.258$\textbf{ or }$v<c_6$ \textbf{then}\\
\quad\quad\quad\textbf{if } $v<c_6-\log\{1+1/(2\kappa)\}$\textbf{ or }$v<-\log\{I_0(\kappa)\}$ \textbf{then}\\
\quad\quad\quad\quad\quad\Return{$\kappa$}
\EndLoop
\end{algorithmic}
\caption{Optimized loop replacing lines \ref{algline:startloop}--\ref{algline:returnline} of algorithm~\ref{alg:kappa}}
\label{alg:squeezed}
\end{algorithm}

\section{Efficiency analysis}
\label{sect:efficiency}
We now analyze the efficiency of Algorithm~\ref{alg:kappa}.  When using the Winitzki approximation, the initial setup (lines  \ref{algline:startsetup}--\ref{algline:endsetup}) involve arithmetic operations, four square roots, two Bessel function evaluations, one logarithm and one exponentiation. In total this setup requires approximately 70 microseconds on a 2.4GHz Intel i5 computer when using the R package \textbf{gsl}. Implementation in a lower-level language would increase the speed significantly.

Each iteration of the rejection loop requires a gamma sample, a uniform sample, between three and five logarithms, and in the worst case a Bessel function evaluation.  The squeeze in Algorithm~\ref{alg:squeezed} does a good job of avoiding the Bessel computation most of the time, and each iteration of the loop requires approximately $10$ microseconds.  Most of these iterations are accepted, and the algorithm, implemented in R, yields approximately 80,000 von Mises samples per second when $\eta=10$ and $\beta_0$ is drawn uniformly over $(-1,1)$.  When using a compiled language such as C++, the algorithm yields over one million samples per second.

\begin{figure}[htp]
\center
\begin{tikzpicture}
\begin{axis}[width=12cm,height=6cm, xlabel=$\beta_0$, ylabel={Efficiency, $\eta=1$}]
\addplot[color=red,dashed,very thick] table[x index=0,y index=1] {Eff1.csv};
\addplot[color=black,very thick] table[x index=0,y index=2] {Eff1.csv};
\end{axis}
\end{tikzpicture}
\begin{tikzpicture}
\begin{axis}[width=12cm,height=6cm, xlabel=$\beta_0$, ylabel={Efficiency, $\eta=10$}]
\addplot[color=red,dashed,very thick] table[x index=0,y index=1] {Eff10.csv};
\addplot[color=black,very thick] table[x index=0,y index=2] {Eff10.csv};
\end{axis}
\end{tikzpicture}
\begin{tikzpicture}
\begin{axis}[width=12cm,height=6cm, xlabel=$\beta_0$, ylabel={Efficiency, $\eta=100$}]
\addplot[color=red,dashed,very thick] table[x index=0,y index=1] {Eff100.csv};
\addplot[color=black,very thick] table[x index=0,y index=2] {Eff100.csv};
\end{axis}
\end{tikzpicture}
\caption{Expected probability of accepting a proposed $\kappa$ (eq.~\ref{eq:acceptprob}), for $\eta=1,10,100$.  Red dashed lines correspond to algorithm~\ref{alg:kappa} with the Winitzki approximation, black lines correspond to the numerically computed true optimal values for $\beta$ and $\kappa_0$.}
\label{fig:Efficiency}
\end{figure}
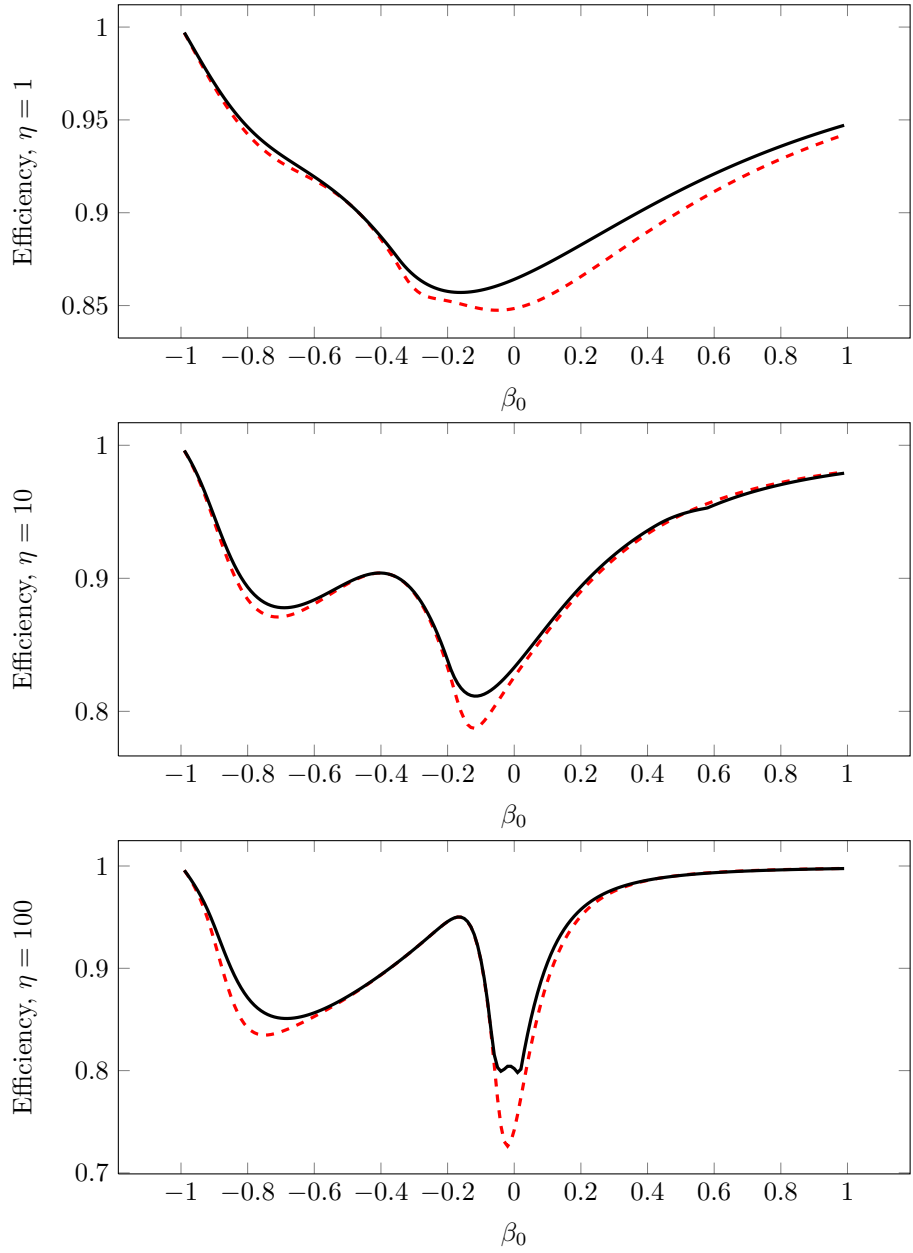

In Figure~\ref{fig:Efficiency} we plot the expected probabilities of acceptance as functions of $\eta$ and $\beta_0$.  The figures were generated by numerically integrating the expected probability of acceptance \eqref{eq:acceptprob} for each $\eta=1,5,10,100$  and for each of 2000 equally spaced  values of $\beta_0\in(-1,1)$. 

There is a noticeable dip in efficiency near $\beta_0=0$. Recalling that $\beta_0=-n^{-1}\sum_{i=1}^n \cos(\theta_i-\mu)$, we see that this region corresponds to diffused $\theta_i$, i.e. the true $\kappa$ is near zero.  This is precisely the region where our Bessel function approximation fails, so this drop is to be expected. Fortunately the drop in efficiency is not severe and our efficiency remains above $0.7$ for all $\eta$ and $\beta_0$.

From Figure~\ref{fig:Efficiency} we see that our algorithm with the approximate optimal parameters does noticeably worse than the numerically computed true optimal parameters when $\beta_0\approx 0$.  This corresponds to the transition region where the optimal $\beta$ rapidly drops from its upper limit of $\beta_0+1$ to its lower limit of $\beta_0+r(\kappa_0)$. Our approximation of the optimal $\beta$ is inaccurate in this transition region. A more sophisticated approximation of the optimal $\beta$ would increase the algorithm's efficiency, however, the region $\kappa\approx 0$ is not usually an area of primary interest and we prefer to use the faster approximation. 

\section{Conclusions}

We have described a highly efficient algorithm to sample from the Bessel exponential distribution.  It is suitable for any application where one wishes generate samples from the posterior distribution for the concentration parameter of the von Mises distribution.

\setlength{\baselineskip}{10pt plus 2pt minus 2pt}

\begin{thebibliography}{}

\bibitem[\protect\citeauthoryear{Abramowitz and Stegun}{Abramowitz and
  Stegun}{1964}]{specialfcnshandbook}
Abramowitz, M. and I.~Stegun (1964).
\newblock {\em Handbook of Mathematical Functions with Formulas, Graphs, and
  Mathematical Tables}.
\newblock Government Printing Office, Washington D.C.

\bibitem[\protect\citeauthoryear{Amos}{Amos}{1974}]{amos}
Amos, D. (1974).
\newblock Computation of modified {B}essel functions and their ratios.
\newblock {\em Mathematics of Computation\/}~{\em 28\/}(125).

\bibitem[\protect\citeauthoryear{Best and Fisher}{Best and
  Fisher}{1979}]{bestfisher}
Best, D.~J. and N.~I. Fisher (1979).
\newblock Efficient simulation of the von {M}ises distribution.
\newblock {\em Journal of the Royal Statistical Society Series C\/}~{\em
  28\/}(2), 152--157.

\bibitem[\protect\citeauthoryear{Boomsma, Mardia, Taylor, Ferkinghoff-Borg,
  Krogh, and Hamelryck}{Boomsma et~al.}{2008}]{boomsma}
Boomsma, W., K.~V. Mardia, C.~C. Taylor, J.~Ferkinghoff-Borg, A.~Krogh, and
  T.~Hamelryck (2008).
\newblock A generative, probabilistic model of local protein structure.
\newblock {\em Proceedings of the National Academy of Sciences\/}~{\em 105},
  8932--8937.

\bibitem[\protect\citeauthoryear{Dagpunar}{Dagpunar}{1978}]{dagpunar}
Dagpunar, J. (1978).
\newblock Sampling of variates from a truncated gamma distribution.
\newblock {\em Journal of Statistical Computation and Simuation\/}~{\em 8},
  59--64.

\bibitem[\protect\citeauthoryear{Damien and Walker}{Damien and
  Walker}{1999}]{damien}
Damien, P. and S.~Walker (1999).
\newblock A full {B}ayesian analysis of circular data using the von {M}ises
  distribution.
\newblock {\em The Canadian Journal of Statistics\/}~{\em 27\/}(2), 291--298.

\bibitem[\protect\citeauthoryear{Forbes and Lauritzen}{Forbes and
  Lauritzen}{2013}]{forbeslasr}
Forbes, P. G.~M. and S.~Lauritzen (2013).
\newblock Fingerprint analysis using {B}ayesian alignment.
\newblock In {\em Proceedings of the {L}eeds Annual Statistics Research
  Workshop}. Available:
  \url{http://www1.maths.leeds.ac.uk/statistics/workshop/lasr2013/proceedings/Forbes.pdf}.

\bibitem[\protect\citeauthoryear{Frellsen, Moltke, Thiim, Mardia,
  Ferkinghoff-Borg, and Hamelryck}{Frellsen et~al.}{2009}]{frellsen}
Frellsen, J., I.~Moltke, M.~Thiim, K.~V. Mardia, J.~Ferkinghoff-Borg, and
  T.~Hamelryck (2009).
\newblock A probabilistic model of local {RNA 3-D} structure.
\newblock {\em Public Library of Science Computational Biology\/}~{\em 5},
  1--11.

\bibitem[\protect\citeauthoryear{Galassi, Davies, Theiler, Gough, and
  Jungman}{Galassi et~al.}{2009}]{gsl}
Galassi, M., J.~Davies, J.~Theiler, B.~Gough, and G.~Jungman (2009).
\newblock {\em {GNU} Scientific Library Reference Manual, Third Edition}.
\newblock Network Theory Ltd.

\bibitem[\protect\citeauthoryear{Guttorp and Lockhart}{Guttorp and
  Lockhart}{1988}]{guttorp}
Guttorp, P. and R.~A. Lockhart (1988).
\newblock Finding the location of a signal: A {B}ayesian analysis.
\newblock {\em Journal of the American Statistical Association\/}~{\em
  83\/}(402), 322--330.

\bibitem[\protect\citeauthoryear{Mardia}{Mardia}{2007}]{mardialasr}
Mardia, K.~V. (2007).
\newblock On some recent advancements in applied shape analysis and directional
  statistics.
\newblock In {\em Proceedings of the {L}eeds Annual Statistics Research
  Workshop}. Available:
  \url{https://www1.maths.leeds.ac.uk/statistics/workshop/lasr2007/proceedings/mardia.pdf}.

\bibitem[\protect\citeauthoryear{Mardia}{Mardia}{2013}]{mardia2013}
Mardia, K.~V. (2013).
\newblock Statistical approaches to three key challenges in protein structural
  bioinformatics.
\newblock {\em Journal of the Royal Statistical Society Series C\/}~{\em 62},
  487--514.

\bibitem[\protect\citeauthoryear{Mardia and Jupp}{Mardia and
  Jupp}{1999}]{mardiabook}
Mardia, K.~V. and P.~E. Jupp (1999).
\newblock {\em Directional Statistics\/} (2nd ed.).
\newblock Chichester, UK: Wiley.

\bibitem[\protect\citeauthoryear{Marsaglia}{Marsaglia}{1984}]{exactApprox}
Marsaglia, G. (1984).
\newblock The exact-approximation method for generating random variables in a
  computer.
\newblock {\em Journal of the American Statistical Association\/}~{\em
  79\/}(385), 218--221.

\bibitem[\protect\citeauthoryear{Marsaglia and Tsang}{Marsaglia and
  Tsang}{2000}]{gammasampler}
Marsaglia, G. and W.~W. Tsang (2000).
\newblock A simple method for generating gamma variables.
\newblock {\em ACM Transactions on Mathematical Software\/}~{\em 26\/}(3),
  363--372.

\bibitem[\protect\citeauthoryear{Winitzki}{Winitzki}{2003}]{winitzki}
Winitzki, S. (2003).
\newblock Uniform approximations for transcendental functions.
\newblock In {\em Computational Science and Its Applications}, Volume 2667 of
  {\em Lecture Notes in Computer Science}, pp.\  780--789. Springer.

\end{thebibliography}

\end{document}